**Superconductivity up to 29 K in SrFe$_2$As$_2$ and BaFe$_2$As$_2$ at high pressures**

**Patricia L. Alireza, Y. T. Chris Ko, Jack Gillett, Chiara M. Petrone, Jacqueline M. Cole, Suchitra E. Sebastian[1], and Gilbert G. Lonzarich**

Department of Physics, University of Cambridge, Cavendish Laboratory, JJ Thomson Avenue, Cambridge CB3 0HE, UK

We report the discovery of superconductivity at high pressure in SrFe$_2$As$_2$ and BaFe$_2$As$_2$. The superconducting transition temperatures are up to 27 K in SrFe$_2$As$_2$ and 29 K in BaFe$_2$As$_2$, making these the highest pressure-induced superconducting materials discovered thus far.

## 1. Introduction

Recently a new class of high-temperature superconductors has been discovered based on iron arsenide layered structures. As in the cuprates and other examples of unconventional superconductivity, the parent compounds tend to be antiferromagnetic and superconductivity emerges under chemical doping or in some cases at high pressure once antiferromagnetism is suppressed.

The magnetic parent compounds of the iron-arsenide family of superconductors crystallize in tetragonal structures, with LaFeAsO [ref. 1] forming in the tetragonal ZrCuSiAs structure, and $A$Fe$_2$As$_2$ ($A$ = Ba, Sr, Ca) forming in the more familiar tetragonal ThCr$_2$Si$_2$ structure (Fig. 1). The spin and charge transitions exhibited by the $A$Fe$_2$As$_2$ materials at ambient pressure are suppressed by hole doping, leading to superconductivity with a transition temperature $T_{sc}$ as high as 38 K [ref. 2]. The application of pressure has also been shown to suppress the spin and charge transitions in these materials [ref. 3, 4, 5]. Here we report high pressure measurements that reveal a superconducting dome in SrFe$_2$As$_2$ and BaFe$_2$As$_2$ with maximum $T_{sc}$ of approximately 30 K. This constitutes the highest pressure-induced observation of superconductivity in any material thus far, to the best of our knowledge.

---

[1] Correspondence to be addressed to S.E.S. (suchitra@phy.cam.ac.uk)

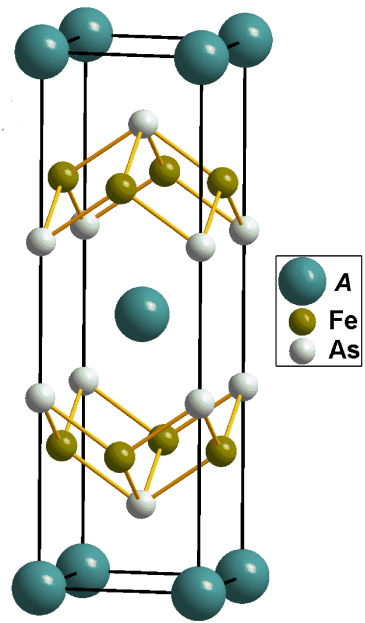

Fig 1: Nominal parent crystal structure of $A$Fe$_2$As$_2$ ($A$ = Ba, Sr, Ca), of the ThCr$_2$Si$_2$ type.

## 2. Experimental details and results

Single crystals of BaFe$_2$As$_2$ and SrFe$_2$As$_2$ were prepared by the flux growth technique [refs. 6, 7] and starting elements of greater than 99.99% purity. SrFe$_2$As$_2$ crystals were grown using Fe:As:Sn flux, and BaFe$_2$As$_2$ using FeAs flux. Crystals were characterised by x-ray diffraction, electron beam microprobe analysis and measurement of the temperature dependences of the magnetic susceptibility and electrical resistivity. Notably, while all single crystals were stoichiometric within limits of resolution of electron probe microanalysis (Sn inclusion in SrFe$_2$As$_2$ single crystals was no higher than 0.3%), not all samples show superconductivity in the reported pressure range. Superconductivity was favoured in SrFe$_2$As$_2$ single crystals grown out of an Fe-rich flux, although there was no detectable deviation from stoichiometry.

Single-crystal X-ray diffraction was used to obtain possible clues as to the sample dependence of superconductivity under pressure. A 300 x 275 x 60μm crystal of SrFe$_2$As$_2$ that showed superconductivity under pressure was mounted onto a Rigaku SCXmini diffractometer, equipped with an Oxford Cryosystems Nitrogen cryostream. Unit cell parameters were determined above and below the Neel temperature, at $T$ = 220 K and 155 K, respectively. Results, given in Table 1, show that the structure is consistent with previous reports where there is much discussion as to the exact nature of its body-centred tetragonal, face-centred orthorhombic or possibly C-centred monoclinic characteristics [refs. 8-11]. Significant diffraction patterns appear in the diffraction patterns – of particular interest is the recurrent characteristic v-shaped diffuse scattering signature throughout reciprocal-space, illustrated in Figure 2. This reveals that two-dimensional disorder is present in the three-dimensional crystal structure in some form. The two most likely origins of this disorder are (i) substantial defects in the *ab* crystallographic plane, or (ii) twinning. Indeed, Table 1 reveals tell-tale signs of a classical form of twinning: the fact that $a \approx b$ in any of the possible options listed, and that $a$ and $b$ in the body-centred tetragonal and face-centred orthorhombic options are related by a factor of $\sqrt{2}$. The exact nature of the three-dimensional crystal structure of $SrFe_2As_2$ with relation to the appearance of superconductivity under pressure is the subject of on-going work.

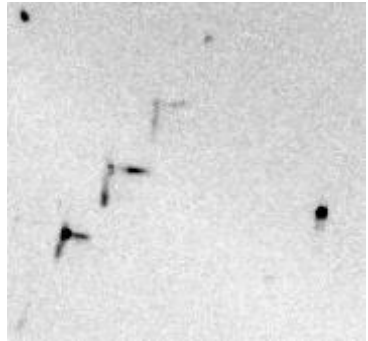

Figure 2: Characteristic diffuse scattering signatures in $SrFe_2As_2$, indicating two-dimensional disorder in the *ab* plane.

Table 1 – Unit cell parameters for alternative structural possibilities of $SrFe_2As_2$ above and below the Neel temperature.

| Bravais Lattice | $a$ / Å | $b$ / Å | $c$ / Å | $\alpha$ /° | $\beta$ /° | $\gamma$ /° | Least-squares fit of data to unit cell |
|---|---|---|---|---|---|---|---|
| $T$ = 220 K | | | | | | | |
| Tetragonal (I) | 3.86 | 3.86 | 12.10 | 90 | 90 | 90 | 2.18 |
| Orthorhombic (I) | 3.84 | 3.97 | 12.16 | 90 | 90 | 90 | 1.51 |
| Orthorhombic (F) | 5.43 | 5.54 | 12.11 | 90 | 90 | 90 | 1.79 |
| Monoclinic (C) | 12.69 | 3.90 | 3.90 | 90 | 106.42 | 90 | 0.72 |
| | | | | | | | |
| $T$ = 155 K | | | | | | | |
| Tetragonal (I) | 3.83 | 3.82 | 12.14 | 90 | 90 | 90 | 0.29 |
| Orthorhombic (I) | 3.81 | 3.84 | 12.14 | 90 | 90 | 90 | 0.09 |
| Orthorhombic (F) | 5.40 | 5.41 | 12.14 | 90 | 90 | 90 | 0.27 |
| Monoclinic (C) | 12.71 | 3.83 | 3.81 | 90 | 107.35 | 90 | 0.09 |

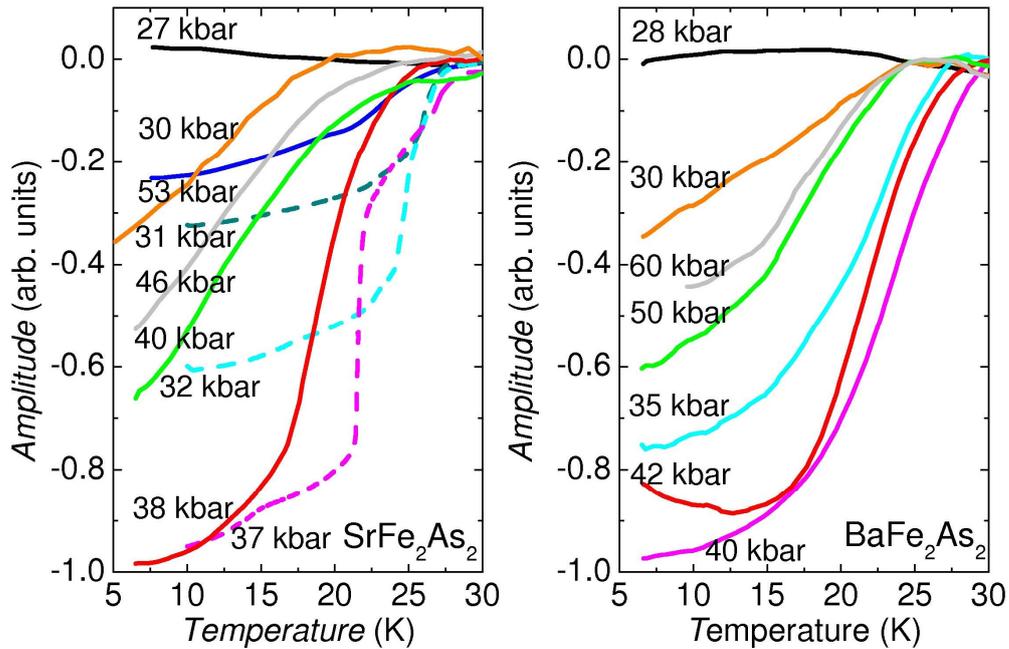

Figure 3: SQUID magnetisation data (solid lines) and AC susceptibility data (dashed lines), measured on plate-like samples approximately 200 μm x 200 μm x 60 μm with the *c*-axis normal to the plate face. Shielding is observed in the pressure range ~ 28 to 60 kbar, consistent with bulk superconductivity within a more narrow pressure range around 40 kbar in both materials. Data shown for (a) $SrFe_2As_2$ and (b) $BaFe_2As_2$, with the magnetic field (50 G in the case of DC magnetisation, and 10 G in the case of AC susceptibility) applied parallel to the *c*-axis. Details of the experimental technique are in ref. 12,13.

Superconducting transitions were detected by means of a miniature diamond anvil cell (the L-A cell – ref. 12) with ultra-low magnetic susceptibility designed for use with a SQUID magnetometer, i.e., the Magnetic Properties Measurement System made by Quantum Design [details of experimental technique are in ref.12]. The pressure transmitting medium is Daphne Oil 7373 and the pressure is measured at room temperature by means of ruby fluorescence before and after each cool down. The change in pressure on cooling to 5 K, the base temperature of our study, has been checked using the known pressure dependence of $T_{sc}$ of a Pb sample and is typically less than 3 kbar in our experiments. A superconducting anomaly is clearly visible upon cooling as a change in the magnetic moment of the sample as a function of decreasing temperature in the presence of an applied magnetic field of 50 G (shown in Fig. 3). Superconducting anomalies have also been measured using an ac field modulation technique with a detection microcoil mounted around the sample inside the sample space of a moissanite anvil cell [details of experimental technique are in ref. 13]. The pressure transmitting medium and the method to determine the pressure is the same as for DC magnetisation measurements. In a narrow pressure range close to 40 kbar, the size of magnetic moment screening below $T_{sc}$ is close to that expected if superconductivity exists throughout the sample (Figures 3 and 4). Bulk superconductivity is inferred by comparing the measured signal with that of a Pb sample of similar size and shape.

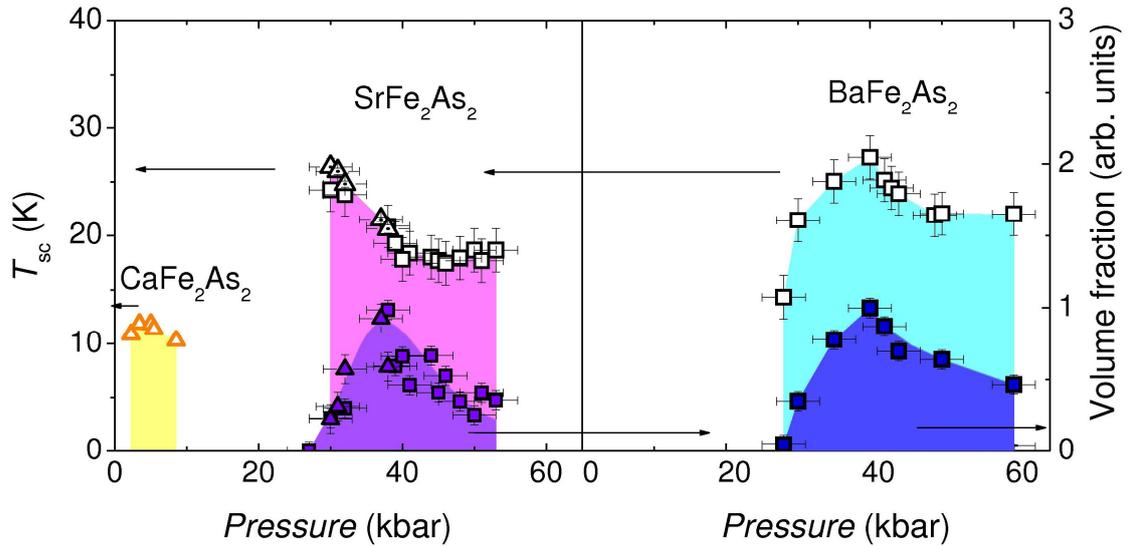

Figure 4: The superconducting transition temperature and superconducting volume fraction of $A$Fe$_2$As$_2$ ($A$ = Sr, Ba) as a function of pressure. The white squares (triangles) show the critical temperature measured by SQUID magnetisation and (AC susceptometry). The filled squares (triangles) show the volume fraction measured by SQUID magnetisation (AC susceptometry). The superconducting dome for CaFe$_2$As$_2$ is taken from ref. 3. SrFe$_2$As$_2$ reveals a maximum critical temperature of ~27 K, while that of BaFe$_2$As$_2$ is ~29 K. Curiously, while superconductivity in BaFe$_2$As$_2$ appears more gradually with increasing pressure, the onset of superconductivity in SrFe$_2$As$_2$ occurs abruptly, accompanied by a maximum in superconducting temperature. Surprising also is the narrow pressure range at which bulk superconductivity is almost complete in both materials.

Figure 4 shows the measured pressure dependences of $T_{sc}$ in BaFe$_2$As$_2$ and SrFe$_2$As$_2$. $T_{sc}$ denotes the temperature at which 25% of the total drop in signal from shielding is reached. We see that the superconducting dome in BaFe$_2$As$_2$ is relatively broad, extending between about 28 kbar to 60 kbar and has a peak of $T_{sc} \cong 29$ K near 40 kbar. In contrast, the superconducting dome is narrower in SrFe$_2$As$_2$, showing a peak of $T_{sc} \cong 27$ K near 30 kbar and a curiously sharp onset around 28 kbar. Interestingly, in both materials, the volume fraction of superconductivity is near complete only over a very narrow pressure range, with the superconducting dome bounded by a decrease in superconducting volume fraction.

## 3. Discussion

An interesting finding is the decrease in the peak of $T_{sc}$ on reducing the ionic size of $A$ in $A$Fe$_2$As$_2$ from Ba and Sr to the isoelectric element Ca (Fig. 4). Also striking is the appreciably broader dome of superconductivity in BaFe$_2$As$_2$ as compared to the Ca and Sr analogues. This increase in the height of the superconducting dome in going from CaFe$_2$As$_2$ to SrFe$_2$As$_2$ and BaFe$_2$As$_2$ may be connected in part with the degree of abruptness with which the spin and charge transitions are suppressed with pressure. Evidence suggests that these transitions are more strongly discontinuous in $CaFe_2As_2$ than in $BaFe_2As_2$ [ref. 14]. The sharper discontinuity may lead effectively to a truncation of the superconducting dome and potentially therefore a reduction in the peak value of $T_{sc}$ in $CaFe_2As_2$ compared with $BaFe_2As_2$. While pressure induced superconductivity is now ubiquitous in various families of materials, the $AFe_2As_2$ class of materials are unique in manifestation of a strongly varying volume fraction of bulk superconductivity within the superconducting dome, while the superconducting temperature remains fairly high. Further experiments will assist in revealing whether the variation in volume fraction reflects either a macroscopic or microscopic inhomogeneity, perhaps reflecting structural twinning effects or coexisting order parameters. The peak superconducting transition temperatures in $BaFe_2As_2$ and $SrFe_2As_2$ are the highest to be induced thus far in a non-superconducting material by the application of pressure. Pressure is thus seen to be a powerful tuning parameter in engineering materials properties.

**Acknowledgements**

These studies would not have been possible without the help of Sam Brown in developing the diamond anvil cells used in this research. We acknowledge Peter Littlewood, Oliver Welzel, and Malte Grosche for discussions, and Swee Goh, Mark Dean, Emily Russell, Chris Lau, and Emma Smith for experimental assistance. This research has been supported by the Cavendish Laboratory, Trinity College, St. Catherine's College, the Isaac Newton Trust, the Croucher Foundation, the Overseas Scholarship, and the Engineering and Physical Sciences Research Council of the UK.